\documentclass[conference]{IEEEtran}
\usepackage{cite}
\usepackage{xcolor}
\usepackage{pifont}

\usepackage{caption}
\usepackage{subcaption}
\usepackage{amsmath, amssymb}
\allowdisplaybreaks
\usepackage{pifont}
\usepackage{booktabs}
\usepackage[utf8]{inputenc}
\usepackage{algorithm}
\usepackage{algpseudocode} 
\usepackage{amsfonts}
\usepackage{graphicx}
\usepackage{xcolor}
\usepackage{bbm}
\usepackage{amsthm}
\usepackage{bm}
\usepackage{multirow}

\title{A Deep Reinforcement Learning Approach for Dynamic Contents Caching in HetNets}
\author{\IEEEauthorblockN{Manyou Ma and Vincent W.S. Wong}
\IEEEauthorblockA{Department of Electrical and Computer Engineering, The University of British Columbia, Vancouver, Canada}
email: \{manyoum, vincentw\}@ece.ubc.ca
}

\begin{document}
\vspace{-1cm}
 \maketitle
\vspace{-1.5cm}
\begin{abstract}
The recent development in Internet of Things necessitates caching of dynamic contents, where new versions of contents become available around-the-clock and thus timely update is required to ensure their relevance. The age of information (AoI) is a performance metric that evaluates the freshness of contents. Existing works on AoI-optimization of cache content update algorithms focus on minimizing the long-term average AoI of all cached contents. Sometimes user requests that need to be served in the future are known in advance and can be stored in user request queues. In this paper, we propose dynamic cache content update scheduling algorithms that exploit the user request queues. We consider a special use case where the trained neural networks (NNs) from deep learning models are being cached in a heterogeneous network. A queue-aware cache content update scheduling algorithm based on Markov decision process (MDP) is developed to minimize the average AoI of the NNs delivered to the users plus the cost related to content updating. By using deep reinforcement learning (DRL), we propose a low complexity suboptimal scheduling algorithm. Simulation results show that, under the same update frequency, our proposed algorithms outperform the periodic cache content update scheme and reduce the average AoI by up to 35\%. 
\end{abstract}
% ===========================================================================
\section{Introduction}
% ===========================================================================
To handle the ever-increasing growth of data traffic, one promising approach is to cache popular contents using a heterogeneous network (HetNet) architecture~\cite{Shanmugam2013FemtoCaching}. In HetNet caching, a macro base station (MBS) and multiple small-cell base stations jointly serve users within a macrocell. These small-cell base stations have storage capacity and can act as content servers (CSs). Previous research has studied different aspects of caching \textit{static contents}, such as predicting future content popularity~\cite{Hou2017Proactive}, content placement strategies~\cite{maddahali2013fundamental}, and scheduling algorithms design~\cite{zhou2016stochatic,  ma2019optimal}. In the aforementioned works, the static contents, such as videos, typically do not change once they have been created and hence only need to be pushed to the CSs once. 

However, with the proliferation of the Internet of Things (IoT) and the mobile edge computing paradigm, billions of IoT devices are expected to be connected to the fifth-generation (5G) and beyond wireless networks. A particular genre of artificial intelligence (AI)-oriented IoT applications is powered by deep learning (DL) algorithms~\cite{lecun2015deep}. DL techniques have been applied ubiquitously in domains such as autonomous driving, natural language processing, and medical diagnosis. Since the training of DL neural networks (NNs) is computation and memory-intensive, general-purpose cloud computing facilities have been developed to train and maintain NNs, using an ever-growing training dataset with new data continuously added into those platforms. Once an NN has been trained, its size is typically small compared to the raw data (\textit{e.g.}, images, videos) collected by the IoT devices. The size of popular pre-trained NNs ranges from 5~MB (SqueezeNet) to 500~MB (VGG11)~\cite{PyTorch2019vision}. Moreover, the implementation of an NN in the deployment stage is less resource-demanding, compared to the training step in DL. Therefore, it is desirable for the IoT devices to download the trained NNs and execute the AI applications using their onboard chips. Tools have been developed for the deployment of DL algorithms on light-weight computational devices, such as smartphones~\cite{tensor2019flow}. In the literature, the DL frameworks, where NNs are trained in a centralized server and later distributed to the users in the system, have already been proposed for wireless communication and robotics applications~\cite{Naparstek2019Deep,sartoretti2019distributed}. In anticipation of the ubiquitous adoption of these DL frameworks, effective algorithms need to be developed to deliver the trained NNs to the system users. 

We postulate that the trained NNs should be treated as \textit{dynamic contents} since we live in a dynamically changing world with the explosive emergence of new information and the DL NNs need to be re-trained using the newly available data to stay adaptive to these new changes. For dynamic contents caching, ensuring the freshness of the contents that are delivered to the IoT applications is of equal importance as satisfying the conventional quality-of-service (QoS) requirements, such as delay and throughput. Due to the massive number of IoT devices connected to the networks, it may not be possible for the IoT devices to download the NNs directly from the cloud computing server. This is because sending all these data packets (with the NNs as payloads) across the cloud through the core network to the radio access network introduces extra delay overhead, and may increase the level of congestion in the core network and the access links. Hence, the aforementioned HetNet architecture can be adopted to tackle the NNs caching problem. To reduce the data traffic in the MBS, recent versions of the NNs can be cached in the CSs. When a user request arrives, a cached NN is sent to the user by the CS at a \textit{target download time} specified by the user. 

To ensure the freshness of the NNs delivered to the IoT user applications,  we propose to use the \textit{age of information} (AoI)~\cite{kaul2012real, Sun_2017_update} of the delivered NNs as a metric to evaluate the system performance. The AoI of a file depicts the amount of time that has elapsed since the current version of a file is generated. Hence, a smaller AoI corresponds to a file that is more recent. Modelling and optimizing the AoI of a system have attracted much research interest. In~\cite{Yates2017age},  Yates \textit{et al.} used AoI as a metric to evaluate the performance of a caching network, where stochastic arrivals of user requests following a renewal process are assumed, and the long-term average AoI of all the files in the system is minimized. However, we conjecture that in practical systems, many user requests may require the NN to be sent at a specific time in future because IoT devices in general submit their request earlier than the expected time that the NN is being used. Therefore, the number of NNs or files that need to be transmitted in the near future are often known ahead of time and can be used to facilitate the scheduling of cache content update. In this paper, we consider the scenario where user requests arrive before their target download time. We employ multiple queues to keep track of user requests for different NNs that need to be served at different target download times. We require each user in the network to submit a request for downloading an NN before the target download time.

In this paper, we investigate the problem of AoI minimization of dynamic contents caching in a HetNet. Compared to previous studies on AoI in the literature, we utilize the information of the user request queues and the target download times to improve the system performance. We consider a scenario where NNs are being cached by the CSs in a HetNet. The algorithm we develop can also be applied to caching other types of dynamic contents.
% Contributions
The contributions of our work are as follows:
\begin{itemize}
    \item We formulate the problem of caching dynamic contents in a HetNet as a Markov decision process (MDP). The objective is to minimize the average AoI of the NNs that are sent to the IoT applications plus the cost related to updating the cached NNs.

    \item We train a deep Q-network (DQN)~\cite{mnih2015human}-based deep reinforcement learning (DRL) agent to learn the state-action values of the formulated MDPs, and thus reduce the memory space required to store the optimal policy.

     \item We perform simulations and show that compared to the existing strategies that do not utilize the user request queues, such as the periodic update approach, our proposed queue-aware cache content update scheduling algorithms obtain up to a 35\% decrease in the average AoI of the dynamic contents delivered to the users. 
\end{itemize}

The rest of this paper is organized as follows. The system model and the MDP problem formulation are presented in Section~\ref{sec:Chapter2}, where methods for obtaining the optimal solution of the MDP problem are introduced. In Section~\ref{sec:Chapter3}, we propose a DQN-based suboptimal algorithm that solve the formulated problem. Performance evaluation and comparison are presented in Section~\ref{sec:sim}. Section~\ref{sec:con} concludes the paper. 
% ===========================================================================
% Section II: 
% ===========================================================================

\section{System Model and Problem Formulation}\label{sec:Chapter2}
\begin{figure}[t]
    \centering
        \includegraphics[width=0.5\textwidth]{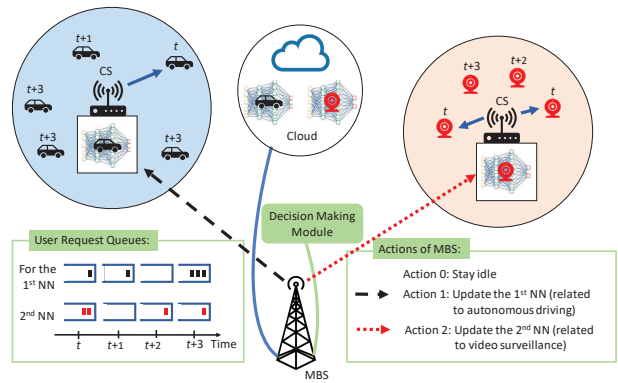}
        \vspace{-0.5cm}
        \caption{System model of the HetNet with one MBS, two CSs, and two NNs. Two different NNs are being cached, one in each CS. One of the NNs corresponds to a navigation system, and the other NN corresponds to computer vision applications. The IoT devices correspond to the IoT-enabled cars and video surveillance cameras in the network. The decision making module is located at the MBS. At a given time instance $t$, the target download time specified by each IoT device is shown above the IoT devices in the figure. The user request queues, which are stored in the decision making module, at time instances $t$, $t+1$, $t+2$, and $t+3$ are shown. The decision making module chooses one of the $(F+1)$ available actions, where action 0 corresponds to staying idle and action $f$ corresponds to updating the $f$-th NN, for all $f\in \mathcal{F}$.}
        \vspace{-0.5cm}
    \label{fig:Caching_neuralNet_two}
\end{figure}

We consider a HetNet consisting of one MBS and $F$ CSs. For the $f$-th CS, where $f \in \mathcal{F} = \{1, \ldots, F\}$, there are $N_f$ users associated with it and an NN  is being cached. We assume only one NN is cached in each CS both for notation simplicity and to ensure that all the CSs can operate simultaneously to serve user requests\footnote{The model can be extended to the cases where (a) multiple NNs are being cached in each CS and (b) each NN is being cached in multiple CSs. For case (a), spectral resources need to be allocated to each CS to ensure the user requests for different NNs can be served simultaneously. For case (b), an NN cached in multiple CSs can be updated via multicasting.}. A sample system topology of the network with two CSs and two different NNs being cached is shown in Fig.~\ref{fig:Caching_neuralNet_two}. In this example, one of the NNs corresponds to the navigation system for IoT-enabled cars and the other NN corresponds to computer vision-based applications for reporting suspicious activities. 

We consider a time-slotted system, and user requests may arrive at the beginning of each time slot. The CSs transmit the latest available cached NN to the users at the beginning of the target download time via multicasting. We assume that the transmission of an NN can be finished within one time slot\footnote{In the case when the NN is large and the transmission cannot be completed within one time slot, the NN training algorithm, which is executed by the cloud computing server, will only update and transmit a subset of the parameters in the NN while the other parameters remain fixed. This approach is known as transfer learning~\cite{Yosinski2014How}.}, and error-free transmission can be achieved\footnote{To consider the possibility of transmission errors, one can extend the state space by including the channel state information.}.  
We consider the use case where multiple AI-enabled IoT devices perform tasks based on NNs stored in their on-board chips. The NNs cached in these devices need to be updated periodically and every time prior to their activation. The NNs are trained in the cloud, and we assume that newer versions of the NNs become available in every time slot. For example, consider a stochastic gradient descent~\cite{kingma2014adam}--based training algorithm running in the cloud with new data added into it after each training epoch. The users can obtain an up-to-date NN after each training step. This is known as the \textit{generate-at-will} model~\cite{Sun_2017_update} in the AoI literature.

The CSs are connected to the MBS via a wireless backhaul. The CSs have disjoint coverage areas. Simultaneous transmissions by different CSs can be achieved when appropriate frequency reuse schemes are adopted. The channels or subcarriers used by the MBS to update the cached NNs are orthogonal to those used by the CSs to serve user requests. At the beginning of a time slot, the CSs serve the user requests that are due at the end of the current time slot. The decision making module in the MBS decides whether the MBS remains idle or updates one of the NNs cached in a CS (see Fig.~\ref{fig:Caching_neuralNet_two}). We assume that apart from updating the NNs cached in the CSs, the MBS also performs other tasks, such as collecting and forwarding data collected from the IoT devices. Therefore, there is a cost associated with allocating time slots for content updating. We assume the cost is linearly dependent on the updating frequency (\textit{i.e.}, average number of updates per time slot), with a coefficient $\eta$.

The MBS is connected to the cloud via a high-speed wired backhaul link. The IoT devices, \textit{e.g.}, the AI-enabled cars shown in Fig.~\ref{fig:Caching_neuralNet_two}, submit their requests to the CS, which are then forwarded to the MBS, for the latest version of NNs at least $\Delta$ time slots before the NN is required. That is, a request submitted at the $t$-th time slot needs to be served in the $(t+\Delta)$-th time slot. For example, $\Delta$ can be the number of time slots that is required for the engine and other hardware in the car to become ready for using the updated NN to perform navigation. The car submits the request when it is turned on at time slot $t$, and an up-to-date NN is delivered when the car is ready to be driven at time slot $t+\Delta$. Since the car will not be driven until $t+\Delta$, sending the NN too early, for example in time slot $t+1$, will lead to a larger AoI of the NN, when the car is actually ready for driving at time slot $t+\Delta$. In this paper, we consider a simple case where all users send the request $\Delta$ time slots before the NNs are needed\footnote{In practice, the system only specifies a minimum time interval $\Delta_{\min}$, which represents the minimum time window between the time a user request is received and the NN is needed. An IoT device may submit a request for an NN which is due well ahead in the future. }. We assume that user request arrivals in the $f$-th CS follow the binomial distribution with rate $\lambda_{f}$. 

\subsection{MDP Problem Formulation}
\label{subsec:CMDP_Formulation}
\subsubsection{Decision Epochs and States}
We consider an infinite horizon MDP, where the decision epochs are represented by the time slots in set $\mathcal{T} = \{0, 1, 2, \ldots \}$. In decision epoch $t \in \mathcal{T}$, let $A^f_t$ denote the AoI of the $f$-th NN being cached in the HetNet. To obtain a finite discrete state space, let $\hat{A}$ denote the upper limit of the AoI. Hence, $\mathcal{A} = \{1,2, \ldots, \hat{A}\}$ is the set of all the possible values of AoI of a cached NN.  In this way, we have $A_t^f \in \mathcal{A}$, $\forall$ $f\in \mathcal{F}$ and $t \in \mathcal{T}$.

In decision epoch $t$, let $Q_t^{f, \delta} \in \mathcal{N}_f\overset{\Delta}{=}\{0, 1, \ldots, N_f\}$ denote the number of user requests for the $f$-th NN that have their target download time at $t+\delta$, where $f\in \mathcal{F}$ and $\delta \in \{0, \ldots, \Delta-1 \}$. Since in the networking literature, \textit{queues} are usually used to denote the number of user requests that have arrived and need to be served, we will refer to $Q_t^{f,\delta}$ as a \textit{user request queue} in this paper. We refer to $\Delta$ as the \textit{window size}. For example, given the window size $\Delta = 3$, at decision epoch $t$, the user request queues that need to be served in decision epochs $t$, $t+1$, and $t+2$ are known. In decision epoch $t$, let $G^f_t \in \mathcal{N}_f$ denote the user request queue of newly arrived user requests for the $f$-th NN, which need to be served in decision epoch $t+\Delta$. 

\begin{figure}[t]
    \centering
    \includegraphics[width=0.5\textwidth]{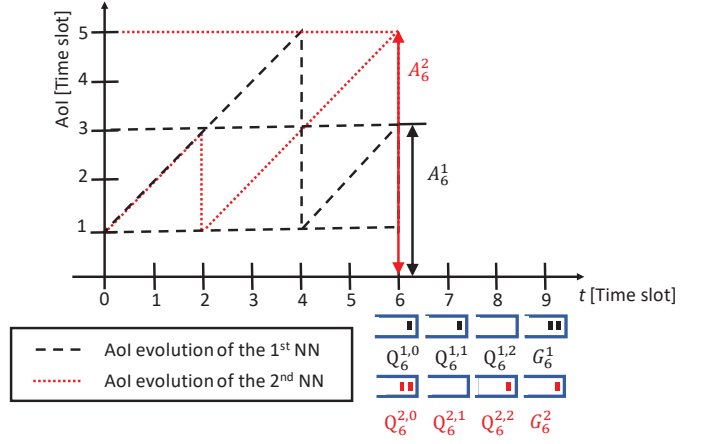}
        \caption{Illustration of the system state in time slot $t=6$ and window size $\Delta = 3$, where $\mathbf{s}_6^1 = (A_6^1, Q_6^{1,0}, Q_6^{1,1}, Q_6^{1, 2}, G_6^1) = (3, 1, 1, 0, 2)$ and states $\mathbf{s}_6^2  = (A_6^2, Q_6^{2,0}, Q_6^{2,1}, Q_6^{2, 2}, G_6^2) = (5, 2, 0, 1, 1)$.}
    \label{fig:AoI_demo_file}
\end{figure}

In decision epoch $t$, as shown in Fig.~\ref{fig:AoI_demo_file}, the set of system states for the $f$-th NN can be represented by a finite set $\mathcal{S}_{f} = \mathcal{A} \times \mathcal{N}_f^{\Delta} \times \mathcal{N}_f$. The state vector for the $f$-th NN can be represented as
\begin{align}
\begin{split}
\mathbf{s}^f_t = &\ \big(A^f_t, Q_t^{f, 0}, Q_t^{f, 1}, \ldots, \\
& \quad Q_t^{f, \Delta-1}, G^f_t \big) \in \mathcal{S}_f, \ \ t\in \mathcal{T}, \ f \in \mathcal{F}.
\label{eq:per_nn_state}
\end{split}
\end{align}

In summary, when considering all the $F$ NNs being cached in the HetNet, the system state space is the finite set $\mathcal{S} = \mathcal{S}_1 \times \cdots \times \mathcal{S}_F$. The state vector $\mathbf{S}_t \in \mathcal{S}$, representing the overall state of the system in decision epoch $t$, can be represented as
\begin{align}
\begin{split}
\mathbf{S}_t &=(\mathbf{s}^1_t, \mathbf{s}^2_t, \ldots, \mathbf{s}^F_t), \ t\in \mathcal{T}.
\label{eq:all_nn_state}
\end{split}
\end{align}

\subsubsection{Actions}
Let $\mathcal{U} = \{0, 1, \ldots, F \}$ denote the set of actions that can be chosen by the MBS. Let $u_t$ denote the action chosen in decision epoch $t$, where the MBS stays idle when $u_t = 0$, updates the $u_t$-th NN in the HetNet when $u_t > 0$. 

\subsubsection{State Transition Probability}
Since each individual user request is independent, in decision epoch $t$, the number of new user request arrivals for the $f$-th NN, $G_t^f$, $f \in \mathcal{F}$, follows the binomial distribution
\begin{align}
\begin{split}
\mathbb{P}(G_t^f = i) =\binom{N_f}{i} \lambda_{f}^{i}(1-\lambda_{f})^{N_f-i}, \ \ \ \ i \in \mathcal{N}_f.
\label{eq:per_nn_P_g}
\end{split}
\end{align}

Given the AoI of the $f$-th NN in decision epoch $t$ and the chosen action $u_t$, its AoI in decision epoch $t+1$ is a deterministic value. We have
\begin{align}
\begin{split}
\mathbb{P}(A^f_{t+1}  \ |\  \mathbf{S}_t, u_t) =  &\ \mathbf{I}(A^f_{t+1} =  A^f_t+1)\mathbf{I}(u_t\neq f)\\
&+\mathbf{I}(A^f_{t+1} = 1)\mathbf{I}(u_t = f),
\label{eq:per_nn_P_A}
\end{split}
\end{align}
\noindent where $\mathbf{I}(\cdot)$ denotes the indicator function. The first case shows that the AoI of the $f$-th NN is increased by 1 if no update is scheduled. On the other hand, if an update is scheduled for the $f$-th NN in decision epoch $t$, then its AoI is reset to 1. 

If a user request arrives in decision epoch $t$, then the target download time is equal to $t+\Delta$. Hence, in decision epoch $t+1$, the value of $Q_{t+1}^{f, \Delta-1}$ depends on whether a new user request arrived in decision epoch $t$. That is,
\begin{align}
\begin{split}
\mathbb{P}(Q^{f, \Delta-1}_{t+1}  \ |\  \mathbf{S}_t) =  \mathbf{I}(Q^{f, \Delta-1}_{t+1} = G^f_t).
\label{eq:per_nn_P_Qend}
\end{split}
\end{align}
\noindent For $Q_{t+1}^{f,\delta}$, $0 \leq \delta \leq \Delta-2$, the state transition probability is given by
\begin{align}
\begin{split}
\mathbb{P}(Q^{f, \delta}_{t+1}  \ |\  \mathbf{S}_t) = \mathbf{I}(Q^{f, \delta}_{t+1} = Q^{f,\delta +1}_t).
\label{eq:per_nn_P_Q}
\end{split}
\end{align}

Given the current state vector $\mathbf{\mathbf{S}}_t$ and action $u_t$, the state transition probability to the next state $\mathbf{\mathbf{S}}_{t+1}$ is equal to
\begin{align}
\begin{split}
\mathbb{P}(\mathbf{S}_{t+1} \ |\ \mathbf{S}_t, u_t ) = & \prod_{f=1}^{F}\bigg(\mathbb{P}(A^f_{t+1} \ |\  \mathbf{S}_t, u_t)\\
&\times\prod_{\delta=0}^{\Delta-1
} \mathbb{P}(Q^{f, \delta}_{t+1}  \ |\  \mathbf{S}_t) \mathbb{P}(G_{t+1}^f)\bigg).
\label{eq:all_nn_P_state}
\end{split}
\end{align}

\subsubsection{Cost}
A deterministic stationary updating policy $\pi$ is defined as a mapping from state space $\mathcal{S}$ to action space $\mathcal{U}$. For a system with state vector $\mathbf{S}_t$, the policy chooses an action $\pi(\mathbf{S}_t) = u_t$, $\forall\  \mathbf{S}_t \in \mathcal{S}$ and $t\in \mathcal{T}$. Similar to the approach in~\cite{zhou2018joint}, we restrict our attention to uni-chain policies, whose induced Markov chain has a single recurrent class (and possibly some transient states)~\cite[vol.~II, Sec. 5.2]{bertsekas2017dynamic1}. In \cite[vol.~II, Proposition~5.2.6]{bertsekas2017dynamic1}, it is stated that for systems satisfying the weak accessibility conditions, there exists an optimal policy that is uni-chain. Since all the system states are reachable with non-zero probability, the weak accessibility conditions hold for our problem. Let 
\begin{align}
\begin{split}
\mathbf{S}^\pi_t = &\  \big(A^{1, \pi}_t, Q_t^{1,0, \pi}, \ldots, Q_t^{1, \Delta-1, \pi}, G^{1,\pi}_t, \ldots,  A^{F, \pi}_t, \\
&\quad Q_t^{F,0, \pi}, \ldots, Q_t^{F, \Delta-1, \pi}, G^{F, \pi}_t\big), \ t\in \mathcal{T},
\end{split}
\end{align}
denote the controlled Markov chain induced by policy $\pi$. Note that $Q_t^{f,0,\pi}$ corresponds to the number of user requests for the $f$-th NN that need to be served in decision epoch $t$. Given policy $\pi$, the expected total AoI of all the served user requests in the first $T$ decision epochs is equal to
\begin{equation}
M_\mathrm{tot}^{\pi}(T) = \sum_{t=0}^{T-1} \sum_{f=1}^{F} \mathbb{E}\left[ A_t^{f, \pi}Q_t^{f,0,\pi} \right],
\label{eq:all_nn_total_aoi}
\end{equation}
\noindent where $\mathbb{E}$ denotes the expectation with respect to the user request arrivals.  The average total number of user requests being served depends on the user request arrival rate and is equal to
\begin{equation}
M_\mathrm{num}(T) = T\sum_{f=1}^{F} N_f\lambda_f.
\label{eq:all_nn_total_arrival}
\end{equation}
Hence, given a policy $\pi$, the average AoI and the update frequency of all user requests can be found as 
\begin{align}
\begin{split}
\overline{M}(\pi)& =  \underset{T \xrightarrow{} \infty}{\lim\sup} \frac{M_\mathrm{tot}^{\pi}(T)}{M_\mathrm{num}(T)},  \\
\overline{C}(\pi) & = \underset{T \xrightarrow{} \infty}{\lim\sup}\ \frac{1}{T} \sum_{t = 0}^{T-1} \mathbb{E} \left[\mathbf{I}(\pi(\mathbf{S}^{\pi}_t) > 0)\right], 
\label{eq:all_average_aoi_cost}
\end{split}
\end{align}
\noindent where $\mathbf{S}^\pi_t$, $\forall\ t\in \mathcal{T}$, follows the state transition probability specified in (\ref{eq:all_nn_P_state}). 

The optimal policy $\pi^*$ is defined to be the uni-chain policy that minimizes the average cost, which corresponds to the average AoI of the served user requests, plus a penalty for each update that is performed. In this case, we can define the cost in each decision epoch $t$ as 
\begin{equation}
c(\mathbf{S}_t, u_t, \eta) \triangleq \frac{\sum_{f =1}^F A_t^{f} Q_t^{f,0} }{\sum_{f =1}^F N_f\lambda_f}+  \eta  \mathbf{I}(u_t > 0).
\label{eq:all_nn_lagrangian_pertime}
\end{equation}
The objective is to minimize the average cost, which comprises the average AoI plus the average updating cost,
\begin{align}
\begin{split}
\overline{L}(\eta, \pi) &= \overline{M}(\pi) + \eta \overline{C}(\pi)\\
&= \underset{T \rightarrow \infty}{\lim \sup} \frac{1}{T} \sum_{t=0}^{T-1}\mathbb{E}\left[ c(\mathbf{S}_t^\pi, \pi(\mathbf{S}_t^\pi), \eta)\right].
\label{eq:all_nn_lagrangian_new}
\end{split}
\end{align}
Problem (\ref{eq:all_nn_lagrangian_new}) is an infinite horizon average-cost MDP problem. Finding its optimal solution involves solving the Bellman equations iteratively, using methods such as the relative value iteration algorithm (RVIA)~\cite[vol. II, Section 5.3.1]{bertsekas2017dynamic1}.

% ===========================================================================
% Section III: Algorithm Design
% ===========================================================================
\section{Algorithm Design}\label{sec:Chapter3}
Since the RVIA algorithm has a large computational complexity, and that the storage space required for storing the optimal policy may be too large for implementation in practical systems, we propose to estimate the state-action value function $V_t(\mathbf{S}_t, u_t, \eta)$ using DRL. In DRL, the state-action values are approximated by an NN which takes the state of the MDP as input\footnote{In this section, the acronym NN is used to denote both the actual neural network we used to estimate the state-action function in the DQN agent as well as the NNs being cached in the HetNet.}. We adopt a state-of-the-art DRL method called DQN~\cite{mnih2015human} to approximate the state-action value function. To avoid overestimating the state-action value function, two NNs with the same dimensions, a \textit{policy network} and a \textit{target network}, are created and being updated during the training steps. We divide the training process into $N_\text{epi}$ episodes to track the training performance, where each episode contains $T_\text{epi}$ \textit{training time steps}. Therefore, there are in total $N_\text{epi}T_\text{epi}$ training time steps, represented by the set $\mathcal{T}_\text{train} = \{0, 1, \ldots, T_\text{epi}N_\text{epi}\}$. At training step $t \in \mathcal{T}_\text{train}$, we denote the parameters of the policy network as $\bm{\theta}_t$ and the parameters of the target network as $\bm{\theta}^\text{target}_t$, which are originally initialized to values sampled from the uniform distribution and updated at each training time step. 

Accounting for the complexity, training time, and the accuracy in approximation, we design an NN with three hidden layers,  with 64, 32, and 16 nodes, respectively. The rectified linear units (ReLUs) are used as the activation functions. In Monte Carlo reinforcement learning, sampled experience  (\textit{i.e.}, simulated interaction with an environment) is used to estimate the state-action value functions instead of RVIA. Given current state $\mathbf{S}_t$ and decision epoch $t$, we use a simulator to sample the next system state $\mathbf{S}_{t+1}$ in decision epoch $t+1$ and cost $c(\mathbf{S}_t, u_t^f, \eta)$. 

Let $V_{t}(\cdot\ |\ \bm{\theta}^\text{target}_t)$ and $V_{t}(\cdot\ |\ \bm{\theta}_t)$ denote the state-action value function approximated by the target network and policy network with parameters $\bm{\theta}^\text{target}_t$ and $\bm{\theta}_t$ at training time step $t$, respectively. We explore an $\epsilon$-greedy policy to avoid overfitting during the training process~\cite{sutton2017reinforcement}. At training time step $t$, it either chooses the best available action in a given state with probability $(1-\epsilon_t)$ or samples a random action with probability $\epsilon_t$. To encourage the DQN algorithm to spend less time exploring the environment after the DQN is well-trained, we apply an exponentially decaying exploration factor $\epsilon_t$, according to 
\begin{equation}
\epsilon_t = \epsilon_\text{min} + (\epsilon_\text{max} - \epsilon_{\text{min}}) e^{-t/\epsilon_{\text{decay}}},
\label{eq:epsilon_greedy}
\end{equation}
\noindent where $0<\epsilon_\text{min} \leq \epsilon_\text{max}<1$ and $\epsilon_\text{decay} > 1$. At each training time step, the best available action is chosen based on the policy network, according to
\begin{equation}
u_t = \underset{u_t \in \mathcal{U}_F}{\arg \min}\  V_{t}(\mathbf{S}_t, u_t, \eta\ |\ \bm{\theta}_t).
\label{eq:choose_action}
\end{equation}
To remove the degree of correlation among the observed sequence of data and to improve the stability of DRL, we adopt the \textit{experience replay}~\cite{kapturowski2018recurrent} approach, where the \textit{system transition tuples} $(\mathbf{S}_t, u_t, c(\mathbf{S}_t, u_t, \eta), \mathbf{S}_{t+1})$ are stored in the \textit{replay memory} after each training time step $t$. At training time step $t$, a set of $K_\text{batch}$ system transition tuples $\mathcal{K}_t^\text{batch}$ are randomly drawn from the replay memory, and \textit{batch gradient descent}~\cite{mnih2015human} is employed to minimize the sum of the loss functions of all the $K_\text{batch}$ system transition tuples. The loss function of an average-cost MDP is defined as 
\begin{align}
    \begin{split}
       L(\bm{\theta}_{t})\  =\  &\frac{1}{2}\bigg(\min_{u^* \in \mathcal{U}_F} V_{t}(\mathbf{S}_{t+1}, u^*, \eta\ |\ \bm{\theta}_t^\text{target}) \\
    & - \min_{u^* \in \mathcal{U}_F} V_{t}(\mathbf{S}^{\text{ref}}, u^*, \eta\ |\ \bm{\theta}_t^\text{target}) \\
    & - V_{t}(\mathbf{S}_t, u_t, \eta\ |\ \bm{\theta}_{t}) + c(\mathbf{S}_t, u_t, \eta) \bigg)^2,
    \label{eq:DRL_lossfunction}
    \end{split}
\end{align}
\noindent where $\mathbf{S}^{\text{ref}}$ is a fixed state that can be chosen arbitrarily and remains fixed during the entire training process~\cite{bertsekas2017dynamic1}.
A stochastic gradient descent step can be expressed as 
\begin{equation}
\bm{\theta}_{t+1} = \bm{\theta}_t - \beta \nabla_{\bm{\theta}_{t}} L(\bm{\theta}_{t}),
\label{eq:DLR_net_update}
\end{equation}
\noindent where $\beta$ is the learning rate for stochastic gradient descent. 
 To avoid overestimating the state-action value function of the optimal action, the parameters of the target network are updated less frequently compared to the parameters of the policy network. The parameters of the target network are updated every $T_\text{update}$ training time steps, by copying $\bm{\theta}_t$ into $\bm{\theta}_t^\text{target}$. In Algorithm~\ref{alg:DQN}, we list the key steps of the algorithm we used to implement the DQN algorithm.
{\linespread{1}
\begin{algorithm}[t]
\begin{small}
\caption{Deep Q-Network (DQN) Algorithm}\label{alg:DQN}
\begin{algorithmic}[1]
\Statex \textbf{Input:} $\eta$, $\epsilon_{\min}$, $\epsilon_{\max}$, $\epsilon_\text{decay} $, $\beta$, $N_\text{epi}$, $T_\text{epi}$, $T_\text{update}$, and $K_\text{batch}$
\State Initialize the replay memory, DQN network parameter $\bm{\theta}_0$, and the target network parameter $\bm{\theta}^\text{target}_0$
\State Observe the initial state $\mathbf{S}_0$ and select a random action $u_0$
\For{$t \in \mathcal{T}_\text{train}$} 
\State System samples $\mathbf{S}_t$ and $c(\mathbf{S}_t, u_t, \eta)$
\State Save the system transition tuple to the replay memory
\State Calculate $\epsilon_t$ according to (\ref{eq:epsilon_greedy})
\State Sample a standard uniform random variable $\epsilon$
\If{$\epsilon < \epsilon_t$}
\State Randomly select an action $u_t \in \mathcal{U}_F$
\Else
\State Choose action according to (\ref{eq:choose_action})
\EndIf
\State Randomly sample a set of $\mathcal{K}_t^\text{batch}$ system transition tuples from the replay memory
\For{$j \in \mathcal{K}_t^\text{batch}$}
\State Calculate the loss function based on (\ref{eq:DRL_lossfunction})
\EndFor
\State Update policy network parameter $\bm{\theta}_t$ based on (\ref{eq:DLR_net_update})
\If{$t\ \text{mod}\ T_\text{update} = 0$}
\State $\bm{\theta}_t^\text{target} \leftarrow \bm{\theta}_t$
\EndIf
\EndFor
\State $\bm{\theta}^\text{target} \leftarrow \bm{\theta}^\text{target}_t$ and $\bm{\theta} \leftarrow \bm{\theta}_t$
\State \Return $\bm{\theta}^\text{target}$
\end{algorithmic}
\end{small}
\end{algorithm}
}

% ==========================================================================
% Simulation Results 
% ---------------------------------------------------------------------------
\section{Performance Evaluation}~\label{sec:sim}
% ===========================================================================
 In this section, we perform simulation studies to validate the analytical results in the paper. Unless specified otherwise, we set $T = 10,000$, $\hat{A} = 50$, $\Delta =4$ and $N_f = 2$, $\forall f \in \mathcal{F}$. We compare the proposed optimal and suboptimal algorithms with the periodic update heuristic proposed in~\cite{Yates2017age}. The DQN algorithm was implemented using PyTorch~\cite{PyTorch2019vision}, and the parameters used for DQN are as follows: $\epsilon_{\min} = 0$, $\epsilon_{\max} = 0.99$, $\epsilon_\text{decay} = 200$, $\beta = 0.01$, $N_\text{epi} = 200$, $T_\text{epi} = 3,000$, $T_\text{update}= 3,000$, and $K_\text{batch} = 1,000$. In this paper, we consider the simple case where $F = 1$, and will address the more general case where $F>1$ in the journal version \cite{ma2019Age} of this work. Each data point represents the average performance over 30 experiments. 
 
\begin{figure}[t!]
    \centering
    \includegraphics[width=0.48\textwidth]{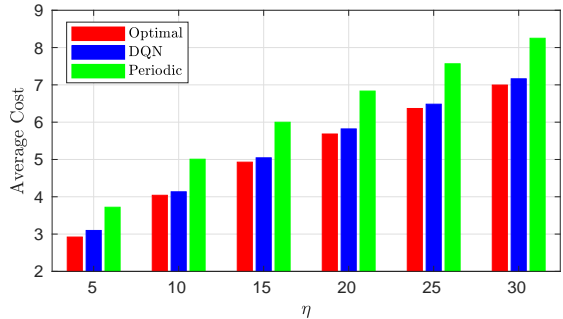}
        \caption{The average cost obtained by the DQN-based approach compared to the periodic update heuristic and the optimal algorithm. $\Delta = 4$ in this case.}
        \vspace{-0.3cm}
    \label{fig:Figure_3}
\end{figure}

\begin{figure}[t!]
    \centering
    \includegraphics[width=0.48\textwidth]{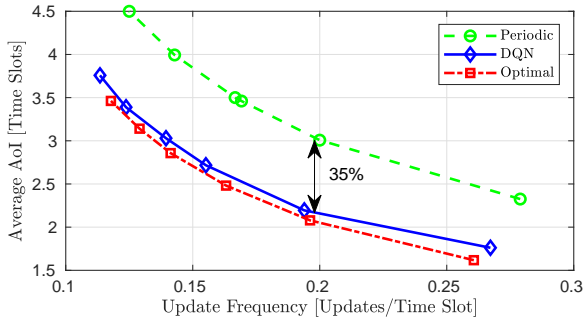}
        \caption{The average AoI obtained by the DQN-based approach compared to the periodic update heuristic and the optimal algorithm v.s. update frequency.}
        \vspace{-0.3cm}
    \label{fig:Figure_4}
\end{figure}

\begin{figure}[t!]
    \centering
    \includegraphics[width=0.48\textwidth]{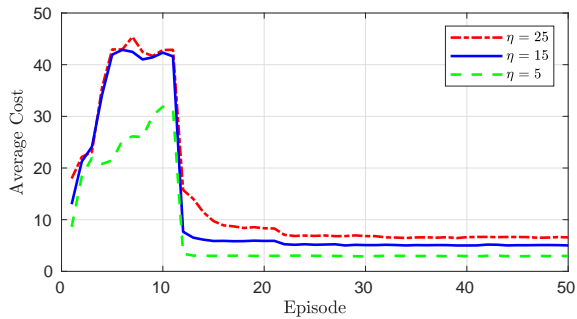}
        \caption{The average cost from subsequent episodes in the DQN training.}
        \vspace{-0.3cm}
    \label{fig:Figure_5}
\end{figure}

 In Fig.~\ref{fig:Figure_3}, we plot the average AoI obtained by the optimal algorithm, the DQN algorithm, and the periodic update heuristic. The periodic update heuristic corresponds to the optimal solution of the formulated MDP problem when $\Delta = 0$ whereas the optimal algorithm is the optimal solution of the formulated MDP when $\Delta = 4$, both of which were found via the RVIA algorithm. We observe that the performance of the DQN algorithm outperforms the periodic update heuristic and is close to the optimal algorithm. To illustrate the advantage of the proposed algorithms, in Fig.~\ref{fig:Figure_4}, we plot the average AoI achieved by the three aforementioned algorithms, against the average updating frequencies of them. The different updating frequencies are achieved by varying $\eta$, where a larger $\eta$ corresponds to updating the NNs less frequently. We observe that under the same updating frequency, the proposed DQN approach can reduce the average AoI up to 35\% and its performance is close to the optimal algorithm found by RVIA.  In Fig.~\ref{fig:Figure_5}, we plot the convergence performance of the DQN algorithm with three different values of $\eta$. We observe the average cost converged in all three cases after around 20 episodes. Therefore, our choice of training the DQN for 200 episodes is appropriate. 

\section{Conclusion}\label{sec:con}
In this paper, we studied the problem of caching dynamic contents using a HetNet architecture. We formulated the problem where the target download time for user requests in a short future time window is known, and designed a strategy where the scheduling decision depends on the user requests that need to be served in the near future. We formulated the problem as an MDP. To reduce the memory required to store the optimal policy of the MDP, we proposed a DRL framework based on DQN to estimate the state-action values of the MDP. Simulation results show that the DQN-based approach has close-to-optimal performance. Both the optimal and suboptimal algorithms outperform the periodic update scheme in different settings. One of the limitations of the DQN approach is that the training time required for good performance is long when multiple NNs are being cached, due to the enlarged state space and action space. We will address this in the journal extension~\cite{ma2019Age} of this work. 
\vspace{0.2cm}
\bibliographystyle{IEEEtran}
\bibliography{IEEEabrv,ICC_bib}

\begin{thebibliography}{10}
\providecommand{\url}[1]{#1}
\csname url@samestyle\endcsname
\providecommand{\newblock}{\relax}
\providecommand{\bibinfo}[2]{#2}
\providecommand{\BIBentrySTDinterwordspacing}{\spaceskip=0pt\relax}
\providecommand{\BIBentryALTinterwordstretchfactor}{4}
\providecommand{\BIBentryALTinterwordspacing}{\spaceskip=\fontdimen2\font plus
\BIBentryALTinterwordstretchfactor\fontdimen3\font minus
  \fontdimen4\font\relax}
\providecommand{\BIBforeignlanguage}[2]{{%
\expandafter\ifx\csname l@#1\endcsname\relax
\typeout{** WARNING: IEEEtran.bst: No hyphenation pattern has been}%
\typeout{** loaded for the language `#1'. Using the pattern for}%
\typeout{** the default language instead.}%
\else
\language=\csname l@#1\endcsname
\fi
#2}}
\providecommand{\BIBdecl}{\relax}
\BIBdecl

\bibitem{Shanmugam2013FemtoCaching}
K.~Shanmugam, N.~Golrezaei, A.~G. Dimakis, A.~F. Molisch, and G.~Caire,
  ``Femto{C}aching: Wireless content delivery through distributed caching
  helpers,'' \emph{{IEEE} Trans. Inf. Theory}, vol.~59, no.~12, pp. 8402--8413,
  Dec. 2013.

\bibitem{Hou2017Proactive}
T.~Hou, G.~Feng, S.~Qin, and W.~Jiang, ``Proactive content caching by
  exploiting transfer learning for mobile edge computing,'' in \emph{Proc. of
  IEEE Global Commun. Conf. (GLOBECOM)}, Singapore, Dec. 2017.

\bibitem{maddahali2013fundamental}
M.~A. Maddah-Ali and U.~Niesen, ``Fundamental limits of caching,'' \emph{{IEEE}
  Trans. Inf. Theory}, vol.~60, no.~5, pp. 2856--2867, May 2014.

\bibitem{zhou2016stochatic}
B.~Zhou, Y.~Cui, and M.~Tao, ``Stochastic content-centric multicast scheduling
  for cache-enabled heterogeneous cellular networks,'' \emph{{IEEE} Trans.
  Wireless Commun.}, vol.~15, no.~9, pp. 6284--6297, Sep. 2016.

\bibitem{ma2019optimal}
M.~Ma and V.~W.~S. Wong, ``An optimal peak hour content server cache update
  scheduling algorithm for 5{G} {HetNets},'' in \emph{Proc. of IEEE Int'l Conf.
  on Commun. (ICC)}, Shanghai, China, May 2019.

\bibitem{lecun2015deep}
Y.~LeCun, Y.~Bengio, and G.~Hinton, ``Deep learning,'' \emph{Nature}, vol. 521,
  no. 7553, pp. 436--444, 2015.

\bibitem{PyTorch2019vision}
\BIBentryALTinterwordspacing
PyTorch. (2019) torchvision.models. [Online]. Available:
  \url{https://pytorch.org/docs/stable/torchvision/models.html}
\BIBentrySTDinterwordspacing

\bibitem{tensor2019flow}
\BIBentryALTinterwordspacing
Google. (2019) Introduction to {T}ensor{F}low lite. [Online]. Available:
  \url{https://www.tensorflow.org/lite/overview}
\BIBentrySTDinterwordspacing

\bibitem{Naparstek2019Deep}
O.~{Naparstek} and K.~{Cohen}, ``Deep multi-user reinforcement learning for
  distributed dynamic spectrum access,'' \emph{{IEEE} Trans. Wireless Commun.},
  vol.~18, no.~1, pp. 310--323, Jan. 2019.

\bibitem{sartoretti2019distributed}
G.~Sartoretti, Y.~Wu, W.~Paivine, T.~S. Kumar, S.~Koenig, and H.~Choset,
  ``Distributed reinforcement learning for multi-robot decentralized collective
  construction,'' in \emph{Distributed Autonomous Robotic Systems}.\hskip 1em
  plus 0.5em minus 0.4em\relax Springer, 2019, pp. 35--49.

\bibitem{kaul2012real}
S.~{Kaul}, R.~{Yates}, and M.~{Gruteser}, ``Real-time status: How often should
  one update?'' in \emph{Proc. IEEE Int'l Conf. on Computer Commun. (INFOCOM)
  Mini-Conf.}, Orlando, FL, Mar 2012.

\bibitem{Sun_2017_update}
Y.~{Sun}, E.~{Uysal-Biyikoglu}, R.~D. {Yates}, C.~E. {Koksal}, and N.~B.
  {Shroff}, ``Update or wait: How to keep your data fresh,'' \emph{{IEEE}
  Trans. Inf. Theory}, vol.~63, no.~11, pp. 7492--7508, Nov. 2017.

\bibitem{Yates2017age}
R.~D. Yates, P.~Ciblat, A.~Yener, and M.~Wigger, ``Age-optimal constrained
  cache updating,'' in \emph{Proc. IEEE Int'l Symp. on Inf. Theory (ISIT)},
  Aachen, Germany, Jun. 2017.

\bibitem{mnih2015human}
V.~Mnih, K.~Kavukcuoglu, D.~Silver, A.~A. Rusu, J.~Veness, M.~G. Bellemare,
  A.~Graves, M.~Riedmiller, A.~K. Fidjeland, G.~Ostrovski \emph{et~al.},
  ``Human-level control through deep reinforcement learning,'' \emph{Nature},
  vol. 518, no. 7540, pp. 529--533, Feb. 2015.

\bibitem{Yosinski2014How}
J.~Yosinski, J.~Clune, Y.~Bengio, and H.~Lipson, ``How transferable are
  features in deep neural networks?'' in \emph{Proc. of Advances in Neural
  Information Processing Systems Conf.}, Montreal, Canada, Dec. 2014.

\bibitem{kingma2014adam}
D.~P. Kingma and J.~Ba, ``Adam: A method for stochastic optimization,'' in
  \emph{Proc. Int'l Conf. on Learning Representations (ICLR)}, San Diego, CA,
  May 2015.

\bibitem{zhou2018joint}
B.~Zhou and W.~Saad, ``Joint status sampling and updating for minimizing age of
  information in the {I}nternet of things,'' \emph{{IEEE} Trans. Commun.},
  vol.~67, no.~11, pp. 7468--7482, Nov. 2019.

\bibitem{bertsekas2017dynamic1}
D.~P. Bertsekas, \emph{Dynamic Programming and Optimal Control, 4th Edition,
  \textit{Vol. I \& II}}.\hskip 1em plus 0.5em minus 0.4em\relax Athena
  Scientific, 2017.

\bibitem{sutton2017reinforcement}
R.~S. Sutton and A.~G. Barto, \emph{Reinforcement Learning: An Introduction,
  2nd Edition}.\hskip 1em plus 0.5em minus 0.4em\relax MIT Press, 2018.

\bibitem{kapturowski2018recurrent}
S.~Kapturowski, G.~Ostrovski, J.~Quan, R.~Munos, and W.~Dabney, ``Recurrent
  experience replay in distributed reinforcement learning,'' in \emph{Proc.
  Int'l Conf. Learn. Representations (ICLR)}, New Orleans, LA, May 2018.

\bibitem{ma2019Age}
M.~Ma and V.~W.~S. {Wong}, ``Age of information driven cache content update
  scheduling for dynamic contents in heterogeneous networks,''
  \emph{\textnormal{submitted to} IEEE Trans. Wireless Commun.}, 2019.

\end{thebibliography}

\end{document}